\documentclass[aps,amssymb,nofootinbib]{revtex4}
\usepackage[T1]{fontenc}
\usepackage[latin9]{inputenc}
\usepackage{color}
\usepackage{amsmath}
\usepackage{amssymb}
\usepackage{graphicx}
\usepackage{esint}
\usepackage{soul}

\makeatletter

\providecommand{\tabularnewline}{\\}

\@ifundefined{textcolor}{}
{%
 \definecolor{BLACK}{gray}{0}
 \definecolor{WHITE}{gray}{1}
 \definecolor{RED}{rgb}{1,0,0}
 \definecolor{GREEN}{rgb}{0,1,0}
 \definecolor{BLUE}{rgb}{0,0,1}
 \definecolor{CYAN}{cmyk}{1,0,0,0}
 \definecolor{MAGENTA}{cmyk}{0,1,0,0}
 \definecolor{YELLOW}{cmyk}{0,0,1,0}
}

\@ifundefined{definecolor}{\@ifundefined{definecolor}
 {\usepackage{color}}{}
}{}
\@ifundefined{definecolor}{\@ifundefined{definecolor}
 {\@ifundefined{definecolor}
 {\usepackage{color}}{}
}{}
}{}

\makeatother

\begin{document}

\title{Effects of modified gravity on B-mode polarization}

\author{{Luca Amendola$^{1}$, Guillermo Ballesteros$^{1}$, Valeria Pettorino$^{1}$}}

\affiliation{$^{1}$Institut f\"{u}r Theoretische Physik, Ruprecht-Karls-Universit\"{a}t
Heidelberg, Philosophenweg 16, 69120 Heidelberg, Germany}

\begin{abstract}
We explore the impact of modified gravity on B-modes, identifying
two main separate effects: lensing and propagation
of tensor modes. The location of
the inflationary peak of the BB spectrum depends on the speed of gravitational
waves; the amplitude of the lensing contribution
depends on the anisotropic stress. We single out these effects using the
quasi-static regime and considering models for which the background
and the growth of matter perturbations are standard. 
Using available data we obtain that the  gravitational wave speed is compatible with the speed of light 
and constrained to within  about 10$\%$.

\end{abstract}

\maketitle

\section{Introduction}

The cosmic microwave background (CMB) is a powerful tool to probe
the early universe and the cosmological evolution that followed. Temperature
fluctuations have been very well measured by independent experiments
\cite{Planck_params2013, 2012arXiv1212.5225B}, including the first cosmological
data released by the Planck collaboration \cite{Ade:2013ktc}. The data are in agreement
with the $\Lambda$CDM model, although still compatible with 
scenarios beyond the standard framework (see for example \cite{Pettorino_2013}). For this reason, it is important to study the potential of new observables than can help to discriminate different models and constrain them further. An obvious possibility are CMB B-modes. Thomson scattering in the presence of primordial fluctuations affects
not only the temperature of the CMB but also its polarization. B-modes generated by gravitational
lensing of the CMB by large scale structure (LSS) have been observed by
two independent teams \cite{Hanson:2013hsb,Ade:2013gez}. In addition,
the BICEP2 collaboration has recently claimed the detection of a B-mode signal
in the CMB (around $\ell\sim80$), interpreted by the
team as the imprint of primordial gravitational waves from inflation
\cite{Ade:2014xna}. Whether this is indeed the case will depend on the analysis
of independent probes, the cross-correlation of data at different
frequencies and a careful check of foregrounds of polarized dust emission
\cite{2012arXiv1212.5225B,Liu:2014mpa,2014arXiv1405.0874P}. 

Cosmological models in which gravity is modified with respect
to Einstein's General Relativity (GR) affect the CMB spectra in several
ways \cite{amendola_etal_2012}, for example through the Late Integrated
Sachs-Wolfe effect \cite{Giannantonio:2009zz} and CMB lensing of
the temperature spectrum \cite{acquaviva_baccigalupi_2006,2004PhRvD..70b3515A}. 
In this paper we study how modified gravity (MG) affects B-mode spectra and 
identify two main observable effects. 

The first effect concerns the lensing
contribution to B-modes and is generated by the anisotropic stress
that characterizes modified gravity theories. Gravitational lensing of the CMB by the large-scale structure (LSS)
affects temperature anisotropies and its ``electric'' (E) and ``magnetic''
(B) polarized components \cite{Lewis_Challinor_2006}. The presence of anisotropic stress in MG  generically  affects
the lensing potential and changes the TT, EE and BB spectra (and the cross-spectra) with respect to those predicted by GR and $\Lambda$CDM.

The second effect is related to the speed $c_{T}$ of gravitational waves, which can
change the peak position of primordial B-modes because $c_T$ determines their epoch of horizon crossing. Once foreground emission
will be under control, these two effects will give a novel handle
for using future B-mode measurements as a tool to test MG. 

There are important difficulties that arise in trying to test
these two effects. First of all, most MG models affect at the same time
the lensing potential and the matter Poisson equation, thereby a change
in the polarization spectra will typically appear associated to a
modification of the growth rate of scalar perturbations. Since the
latter begins to be considerably constrained by the current data (see
e.g. \cite{Macaulay:2013swa}) and the temperature spectrum of scalar
perturbations in the CMB is very well determined \cite{Planck_params},
it could be naively expected that a substantial modification of the
BB-spectrum would also imply a large imprint in the growth
of structures or the TT spectrum. In addition, there is a large
variety of MG models (see for ex. \cite{Amendola2010,Clifton:2011jh,Yoo:2012ug})
and adapting a CMB Boltzmann code for a significant fraction of them
is a daunting task. Conversely, restricting the analysis to a narrow
subset of cases would reduce the appeal of a study of B-modes in MG.

In this paper we propose to bypass these problems by adopting a radical
strategy. We will focus on MG models with a single extra scalar degree
of freedom in the linear quasi-static regime, i.e. taking the large
$k$ limit and neglecting the time derivatives of the potentials.
This reduces considerably the complexity of vast classes of models.
Concretely, this can be applied to the entire class of Horndeski Lagrangians
\cite{Horndeski:1974,Deffayet:2011gz}, which comprises most viable modified gravity models based on a  scalar field, and bimetric gravity models
\cite{Hassan:2011zd}. Furthermore, we assume
that the modification of the Poisson equation is negligible, effectively
selecting those models whose effect on the CMB is essentially due
to lensing and the propagation of gravitational waves. We also assume
that the background behaviour is the one of $\Lambda$CDM. In other words, we assume that MG only affects those observables related to gravitational lensing and gravitational
waves. This choice allows us to test the effects of the anisotropic stress
and the sound speed of gravitational waves with a prescription that can be easily implemented
in Boltzmann codes.

MG effects are in general time dependent. We can therefore test their
impact at various epochs during the history of the universe (e.g.
at decoupling time or at later times). Lensing effects are mostly
generated at a redshift of order unity, while the speed of gravitational
waves affects the CMB at decoupling time and, to a minor extent, at or before
reionization. In the following we will thus treat separately the following
cases: MG effects present at all times, only at decoupling, or
only at low redshift.

\section{Modified gravity in the quasi-static limit}

\label{qsl}

Linear perturbations in MG models are often studied in the
\textit{quasi-static} (QS) limit, obtained for wavenumbers $k$ that
are large compared to the inverse sound horizon, and in which the
additional degrees of freedom with respect to GR (e.g. a scalar field
or a second spin-2 field) do not propagate significantly but rather can be well described
by constraint equations (such as the Poisson equation). The existence
of a valid QS approximation is not always guaranteed and should be
checked on a case by case basis. For instance, it may happen that
the scales at which the limit can be attained are well outside the
linear regime, or it may occur that the validity of the limit depends
on the specific initial conditions for the perturbations. In addition,
the QS limit has to be properly defined in order to avoid spurious
scale dependencies \cite{Bellini2014}.

Working with a perturbed flat FLRW metric in Newtonian gauge 
\begin{equation}
\mathrm{d}s^{2}=-(1+2\Psi)\mathrm{d}t^{2}+a^{2}(1+2\Phi)\mathrm{d}x^{i}\mathrm{d}x^{i}\,,
\label{met}
\end{equation}
the effects of MG in the QS limit can be generally encoded
in two functions \cite{Amendola:2012ky} that reduce to $Y=\eta=1$
in $\Lambda$CDM): 
\begin{equation}
Y(k,a)\equiv-\frac{2k^{2}\Psi}{3{\cal H}^2 \Omega_{m}\delta_{m}}\,,\quad\eta(k,a)\equiv-\frac{\Phi}{\Psi}\,,\label{Y_def}
\end{equation}
where ${\cal H}=aH$ is the conformal Hubble function, $\delta_{m}$ is the matter density contrast and $\Omega_{m}$
is the background matter energy density relative to the critical
one. 
In eq.(\ref{Y_def}) the perturbation variables are meant to denote the standard deviations of the respective quantities; therefore $Y,\eta$ are deterministic functions.
The function $Y$ gives an indication of how the growth
of matter perturbations is altered with respect to the standard one
in GR for large $k$. The function $\eta$ depends
on the two Newtonian gravitational potentials and therefore effectively
on the anisotropic stress. 

It has been shown that $Y$ and $\eta$ take a particularly simple form in broad classes of MG models in which the equations of
motion for the perturbations are of second order (for instance, in the Horndeski Lagrangian or in bimetric gravity)
\cite{DeFelice:2011hq,Amendola:2012ky,2014arXiv1402.1988K,2014arXiv1404.4061S}:
\begin{align}
Y=h_{1}\frac{1+(k/{\cal H})^2  h_{5}}{1+(k/{\cal H})^{2}h_{3}}\,,\quad\eta=h_{2}\frac{1+(k/{\cal H})^2 h_{4}}{1+(k/{\cal H})^{2}h_{5}}\,.\label{limo1}
\end{align}
Here $h_{1-5}$ are functions that depend only on time and can be
obtained directly from the Lagrangian of the model (see Appendix). Several types of Hordenski Lagrangians
that have been studied in detail (see e.g. references in
\cite{Bellini2014}) provide specific examples for which the expressions
(\ref{limo1}) can be applied.

As discussed in the Introduction, we will focus our attention on models
such that $Y=1$ but $\eta\neq1$. With this choice we ensure that
scalar perturbations obey the standard Poisson equation for large
$k$: therefore on scales below the sound horizon, the growth of scalar
perturbations follows the usual one in $\Lambda$CDM. This can occur if
$h_{1}=1$ and $h_{3}\simeq h_{5}$ (see Appendix \ref{appe}). To
simplify our task further, we will also assume that the three remaining
$h$-functions ($h_{2}$, $h_{4}$ and $h_{5}$) can be treated as
constants. This amounts to say that their time variation is slow in
one Hubble time and it is a reasonable approximation for studying
first order scale dependent effects. With this simplification, a very
large class of MG models is mapped into three real constants that
encode the possible effects of the anisotropic stress.
As crude as they may seem, these approximations are a significant
improvement with respect to earlier studies of linear perturbations
in MG, in which $Y$ and $\eta$ were often assumed to be pure constants.

The lensing effect of MG can then be easily described by the deviation
of $\eta$ with respect to unity, which we parametrize as follows:
\begin{align}
1+\eta=2a_{1}\frac{1+a_{2}(k/k_{p})^{2}}{1+a_{3}(k/k_{p})^{2}}\,,\label{def_sigma}
\end{align}
where $a_{1},a_{2},a_{3}$ will be assumed to be constant and we will
take $k_{p}=0.1h$/Mpc.%
\footnote{As usual, $h$ here is the reduced Hubble constant $h=H_{0}/100$,
where $H_{0}$ is the present rate of expansion expressed
in km/s/Mpc.%
}\\

\section{Tensor modes propagation speed}

\label{sec:speed} If the anisotropic stress is non-standard (i.e.
$\eta\not=1$), it can be shown that the tensor modes propagate in
a non-standard way. As discussed earlier, this will modify the CMB
spectra and, in particular, the B--modes. The general form of the
propagation equation for the transverse-traceless amplitude $h$  in  Hordenski Lagrangians can be written (see \cite{DeFelice:2011bh,Bellini2014}
and also \cite{Saltas:2014dha} for a generalization)
as:\footnote{We thank M. Kunz, I. Saltas and I. Sawicki   for discussing with us the structure of this equation.}
\begin{equation}
\ddot{h}+(3+\alpha_{M})H\dot{h}+c_{T}^{2}\frac{k^{2}}{a^{2}}h=0\,\,\,,\,\label{eq:gw}
\end{equation}
where dots represent derivatives with respect to cosmic time and $\alpha_{M},c_{T}$ are functions of time that depend
on the specific model; in the standard case one has $\alpha_{M}=0$
and the gravitational waves speed $c_{T}$ equals the speed of light,
$c_{T}=1$. Notice that the tensor equation is valid in general, not
just in the QS limit. In the notation of \cite{Amendola:2012ky,Bellini2014}
one has: 
\begin{align}
\alpha_{M} & =\dot{w}_{1}/w_{1}H\,\,\,,\\
c_{T}^{2} & =w_{4}/w_{1}\,\,\,,
\end{align}
where $w_{1},w_{4}$ are in general time-dependent functions explicitly
defined in the Appendix that depends on the MG model.  Although in
principle they can be both non-zero, in the specific case we are investigating
here (see Appendix), one has $\alpha_{M}=0$ and 
\begin{equation}
c_{T}^{2}=w_{1}=\frac{1}{2a_1-1}\,\,\,.\label{gwct}
\end{equation}
A decrease of $c_{T}$ moves the horizon crossing of tensor modes
to later times and smaller scales; as a consequence, the BB spectrum
tensor mode first peak moves to higher $\ell$s, as we show later
on. The position of the tensor B peak is therefore a measure of the
gravitational speed at decoupling time.

The speed of gravitational waves can be constrained also with the
gravi-Cherenkov effect (see e.g. \cite{1980AnPhy.125...35C, Moore:2001bv,2012JCAP...07..050K}), which gives an extremely  tight lower limit but no upper limit. Other possible ways to constrain the graviton speed are reviewed in \cite{Goldhaber:2008xy}. However, all these methods apply only locally (or at most within the distance scale of cosmic rays) and/or at the present time; therefore, they are complementary to the observation of B-modes. For other recent analysis on quantum gravity effects see for example \cite{2014arXiv1404.6672C}.

The theoretical BB spectrum shows another peak at $\ell\approx5$,
still to be detected, due to the effects of tensor modes on the scattering
during reionization. Also this peak gets shifted for $c_{T}\not=1$
as we show below. Its detection, for instance with the proposed satellite mission LiteBIRD \footnote{http://litebird.jp/} \cite{2013arXiv1311.2847M}, could therefore put constraints on
the gravitational wave speed before and during reionization.

\section{Weak lensing and CMB spectra}

\label{lensCMB} In order to compute the weak lensing of the CMB by
LSS in a flat universe, we define the lensing potential $\psi$ from
the Weyl potential $\tilde{\Psi}\equiv (\Psi - \Phi)/2$ as follows \cite{Lewis_Challinor_2006} (note that we are using a different signature for the metric in eq. (\ref{met})):
\begin{align}
\psi(\mathbf{\hat{n}})= 2 \int_{0}^{\chi_{\ast}}d\chi\,\tilde{\Psi}(\chi\mathbf{\hat{n}},\tau_{0}-\chi)\,\frac{\chi_{\ast}-\chi}{\chi_{\ast}\chi}\,,\label{lens}
\end{align}
In this expression, $\mathbf{\hat{n}}$ is a unit vector in three-space
that gives the (non-deflected) direction of propagation of a CMB photon,
$\tau_{0}-\chi$ is the conformal time at which the photon was at position
$\chi\mathbf{\hat{n}}$ and $\chi_{\ast}$ is the conformal distance
to the last scattering surface (assuming it is instantaneous). 

The deflection angle with respect to the trajectory that the photon would have in a perfectly
homogeneous universe is given by the angular derivative of the lensing
potential $\alpha = \nabla_{\mathbf{\hat{n}}}\psi(\mathbf{\hat{n}})$. The
lensed CMB temperature $\Theta_{L}$ measured in the direction $\mathbf{\hat{n}}$
corresponds to the unlensed temperature $\Theta$ in the direction
$\mathbf{\hat{n}}+\nabla_{\mathbf{\hat{n}}}\psi$, i.e. 
\begin{align}
\Theta_{L}(\mathbf{\hat{n}})=\Theta\left(\mathbf{\hat{n}}+\nabla_{\mathbf{\hat{n}}}\psi\right)\,.
\end{align}

With the definitions (\ref{Y_def}) we can write 
\begin{equation}
\tilde{\Psi}=\frac{1}{2}(1+\eta)\Psi 
\label{psieq}
\end{equation}
and use this relation to express the lensing potential (\ref{lens})
in terms of $\Psi$ and $\eta$.  We can then write $P_{\tilde{\Psi}} = (1+\eta)^2 P_{\Psi}/4$ where the power spectrum (and similarly the transfer function) of $\tilde{\Psi}$ is written in terms of the one of $\Psi$, which is related to the
matter one by the Poisson equation (\ref{Y_def}).
As noticed before, since the main contribution to CMB lensing comes
from a short range in $z$, peaked around $z\sim1$ \cite{2004PhRvD..70b3515A,acquaviva_baccigalupi_2006,2013JCAP...09..004C, 2013JCAP...02..024A},
the assumption of constant values for $a_{1,2,3}$ is justified. We can then modify a Boltzmann code to include the effect of
MG in the computation of the lensing potential with these approximations.

The power spectrum $C_{l}^{\psi}$ of the lensing potential at a
given $\mathbf{\hat{n}}$ is 
\begin{align}
\langle\psi_{lm}\psi_{l'm'}\rangle=\delta_{ll'}\delta_{mm'}C_{l}^{\psi}\,,
\end{align}
where $\psi_{lm}$ are the coefficients of the expansion of $\psi$
in spherical harmonics: $\psi=\sum_{lm}\psi_{lm}Y_{lm}$. This power
spectrum can be expressed as \cite{Lewis_Challinor_2006}: 
\begin{equation}
C_{\ell}^{\psi}=16\pi\int{\frac{dk}{k}\mathcal{P}_{\mathcal{R}}(k)\left[\int_{0}^{\chi_{\ast}}d\chi\, T_{\tilde{\Psi}}(k;\tau_{0}-\chi)\, j_{l}(k\chi)\,\frac{\chi_{\ast}-\chi}{\chi_{\ast}\chi}\right]^{2}}\,,\label{cll}
\end{equation}
where
${\cal{P}}_{\tilde{\Psi}} = \mathcal{P}_{\mathcal{R}} T_{\tilde{\Psi}}$ and $j_{\ell}(r)=\sqrt{\pi/2r}J_{l+1/2}(r)$, being $J_{l}(r)$ is the $l$--th Bessel function of the first kind. The transfer function
$T_{\tilde{\Psi}}(k;\tau_{0}-\chi)$ propagates the lensing potential (\ref{lens})
forward in time. $\mathcal{P}_{\mathcal{R}}(k)$
denotes the power spectrum of the primordial curvature perturbation
$\mathcal{R}$ at the last scattering surface.
Following eq.(\ref{psieq}), we can then account for the effect of MG by introducing a $(1+\eta)/2$ factor inside the $\chi$
integral of (\ref{cll}). Again, for $\eta = 1$, the integral is the standard one.

Before doing the full calculation with a Boltzmann code, we can get
some insight of the MG effect in the region of relevance for the expected primordial B-mode peak
for large scales $(\ell\ll1000)$ \cite{Lewis_Challinor_2006}. In this limit
and at lowest order in $C_{\ell}^{\psi}$, the lensed B-mode spectrum
$\tilde{C}_{l}^{B}$ is approximately independent of $\ell$ \cite{Lewis_Challinor_2006}: 
\begin{align}
\tilde{C}_{\ell}^{B}\simeq\frac{1}{4}\int d\ell'\ell^{\prime5}C_{\ell'}^{\psi}C_{\ell'}^{E}\,,\label{bmodex}
\end{align}
where $C_{\ell}^{E}$ is the unlensed E-mode spectrum. Since $C_{\ell}^{\psi}$ enters linearly
in (\ref{bmodex}), we see that in modified gravity the lensing contribution
to the B-mode spectrum gets enhanced at large scales by $(1+\eta)^{2}/4$,
with respect to $\Lambda$CDM.

Since we are deriving the lensing MG effects in the QS limit,
i.e. at sub-horizon scales, it is necessary to check the consistency
of our assumption. A comoving mode $k$ translates through the Limber
approximation into a multipole
\begin{equation}
\ell\approx\pi r(z)k\,.
\end{equation}
 One finds that for
a standard $\Lambda$CDM background, $k/{\cal H}>10$ up to $z=7$, and $k/{\cal H}>5$
up to $z=20$, assuming $\ell\ge100$, which is where most of the
BB lensing signal is expected. Since the lensing effect comes from
$z$ of a few at most \cite{Lewis_Challinor_2006}\textbf{, }we expect
the quasi-static approximation to be acceptable.

As we already mentioned, we are neglecting the impact
of MG on the integrated Sachs-Wolfe (ISW) effect. Although $Y=1$
means that there is no effect on the matter perturbation growth, there
will be a change in the ISW due to the fact that photons see both
potentials $\Psi,\Phi$, just as in the lensing case. However, contrary
to lensing, the ISW has a non negligible impact on the temperature
CMB fluctuations at  superhorizon scales, where our quasi-static approximation
is not reliable. We can obtain a rough estimate of the MG effect on
the low-$\ell$ TT spectrum by considering that the ISW contribution
is negligible at $\ell>30$. Then we find that it amounts to less than 20\% (see e.g. \cite{2004A&A...423..811M})
at $\ell>10$, and rises up to 50\% for $\ell=2$. The MG effects
are again proportional to $(1+\eta)^{2}$, although at these
scales the form of $\eta$ is no longer given by the QS expression.
Nevertheless, assuming $a_1\approx1.3$ (see our best fit case below),
we see that the ISW should be increased by 69\%, which means the total
TT spectrum should increase by 14\% at $\ell\approx10$, and up to
35\% at the quadrupole. This is a non negligible effect but it is
well within the cosmic variance. Moreover, it might be possible to absorb
it by slight adjustments of other cosmological parameters,  although we are not
going to explore this possibility  in depth in this paper. 

Let us just briefly point out  what would be the effect of the MG model we consider on the determination of neutrino masses and the total number of relativistic species. As it is well known (see e.g. \cite{Hou:2012xq}), the effect of small neutrino masses on the CMB power spectrum takes place via secondary anisotropies and is multipole dependent. In standard $\Lambda$CDM, a total neutrino mass of the order of $\sum m_\nu\simeq 0.5$ eV reduces the CMB $C_l$ with respect to the case $\sum m_\nu =0$ by at most $\sim 8\%$ for $10< \ell <20$, due to the late time ISW. For larger values of $\ell$, the early time ISW decreases the $C_l$ by smaller amount (around $2\%$ up to $\ell \sim 100$) and then increases it by approximately $1\%$ (for $100<\ell  <500$). As we have just explained the $(1+\eta)^2$ factor introduced in the ISW by the modification  of gravity we consider in this work tends to enhance the $C_l$ with respect to $\Lambda$CDM for all $\ell$. Therefore, for small multipoles ($\ell < 100$) it would go on the opposite direction as that of $\sum m_\nu$, whereas it would enhance the effect of massive neutrinos for $100 < \ell  <500$ (where the early time ISW introduced by $\sum m_\nu$ is rather small). If a $(1+\eta)^2$ factor was present in the CMB, the net overall effect would be an increased ISW that could be approximately compensated by a higher value of $\sum m_\nu$. 

On the other hand, the main CMB effect of a larger number of effective relativistic species $N_{\text{eff}}$ with respect to its standard value of 3.046 is due to a delayed time of matter-radiation equality, which results in a vertical shift of the CMB peaks with respect to the first one \cite{Hou:2011ec}. Concretely, it introduces a shift of the order of $\Delta C_l/C_l\simeq -0.072\, \Delta N_{\text{eff}}$. This means that a $(1+\eta)^2$ ISW change can have a similar effect as that of a reduced number of relativistic species. In conclusion, the kind of MG that we consider can affect the determination of both the total neutrino mass and the number of relativistic species. A detailed study of the degeneracies of these effects could be the object of future work.

In any case, in order not to bias our results, we decided to bypass
this problem in the data analysis below by cutting all the multipoles
$\ell<100$ in the TT and TE spectra. This slightly enlarges our errors
but avoids using the incorrectly modeled low-$\ell$ tail of the temperature
spectrum.
We conclude this paragraph by noting that also weak lensing of the matter power spectrum can be used to test modified gravity models \cite{2013MNRAS.429.2249S} and should be seen as a complementary probe with respect to the lensing on B-mode polarization. On the other hand, we have checked that the corresponding TT and EE spectra are very little affected by the kind of modification of gravity that we study,
%these changes,  
so that 
%in our model 
the main contribution of MG is on B-modes.

\section{Results}

\label{results}

In order to test how modified gravity affects the CMB spectra, we have
modified the publicly available Boltzmann code CAMB.%
\footnote{http://camb.info/%
} Given the assumptions described in \ref{qsl}, we do not need to
modify the background, which is assumed to be $\Lambda$CDM. Within
CAMB, we apply two sets of corrections to perturbations, which can
be activated separately or simultaneously: 
\begin{itemize}
\item we modify the lensing potential as described in eq.~(\ref{psieq}), which
then depends on the $a_{1},a_{2},a_{3}$ parameters; 
\item we modify the tensor propagation equation as described  eq.~(\ref{eq:gw}). This modification only depends
on $a_{1}$.
\end{itemize}
We have included  the three new parameters in COSMOMC \cite{cosmomc_lewis_bridle_2002}.
Both CAMB and COSMOMC used for this paper are the ones from the March
2014 version. After checking that for $a_{1}=1$ and $a_{2}=a_{3}$
we recover standard $\Lambda$CDM, we have then tested separately
three cases: when MG effects are present at all times and therefore
both on lensing and on the gravitational wave speed (we refer to this
case as ``CT+lensing''), when they are present only at decoupling
and therefore only on the tensor speed (``CT''), or only at low
redshift (``lensing'').

\begin{figure}
\centering %
\begin{tabular}{cc}
\includegraphics[width=85mm]{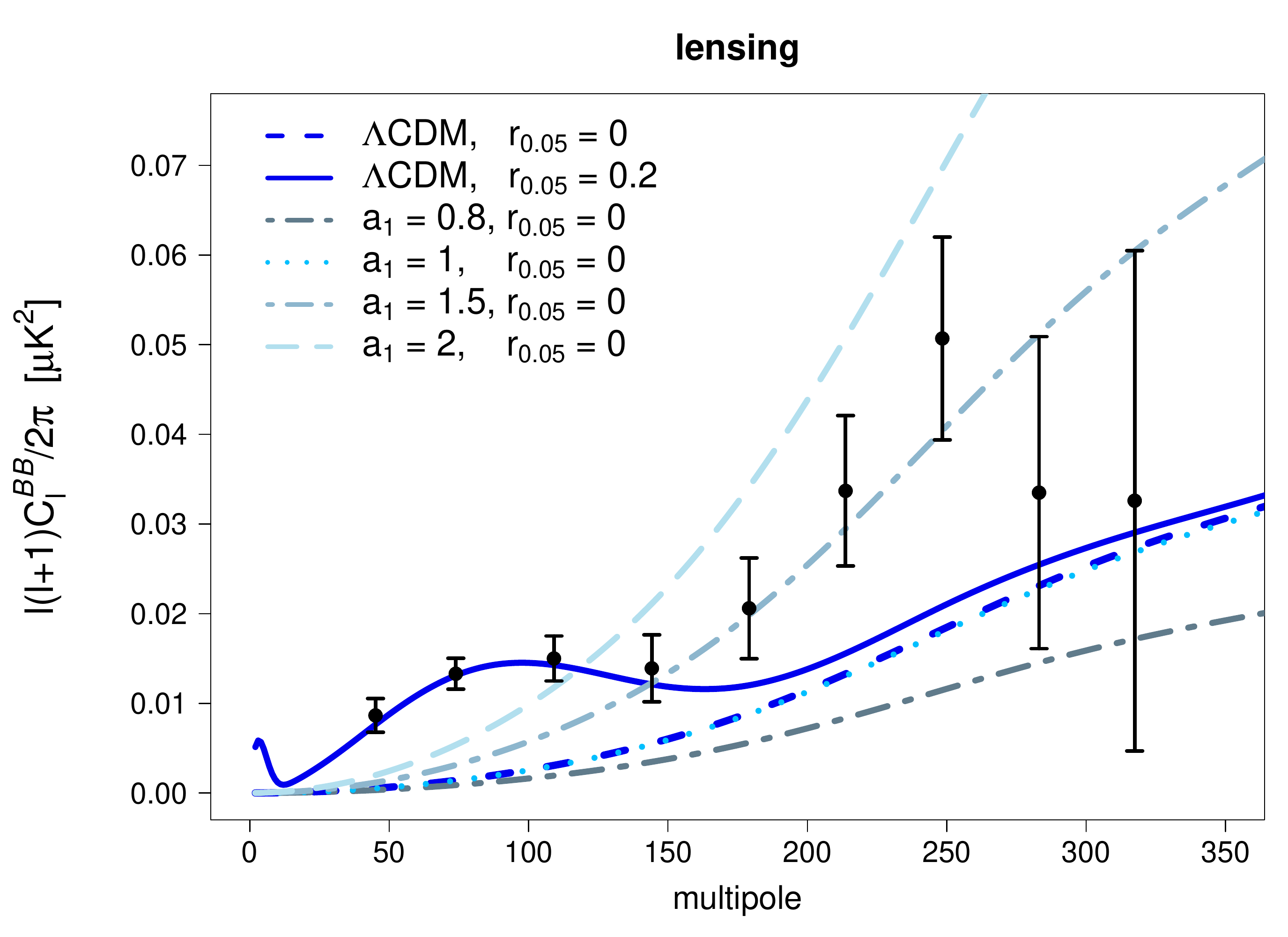}  & \includegraphics[width=85mm]{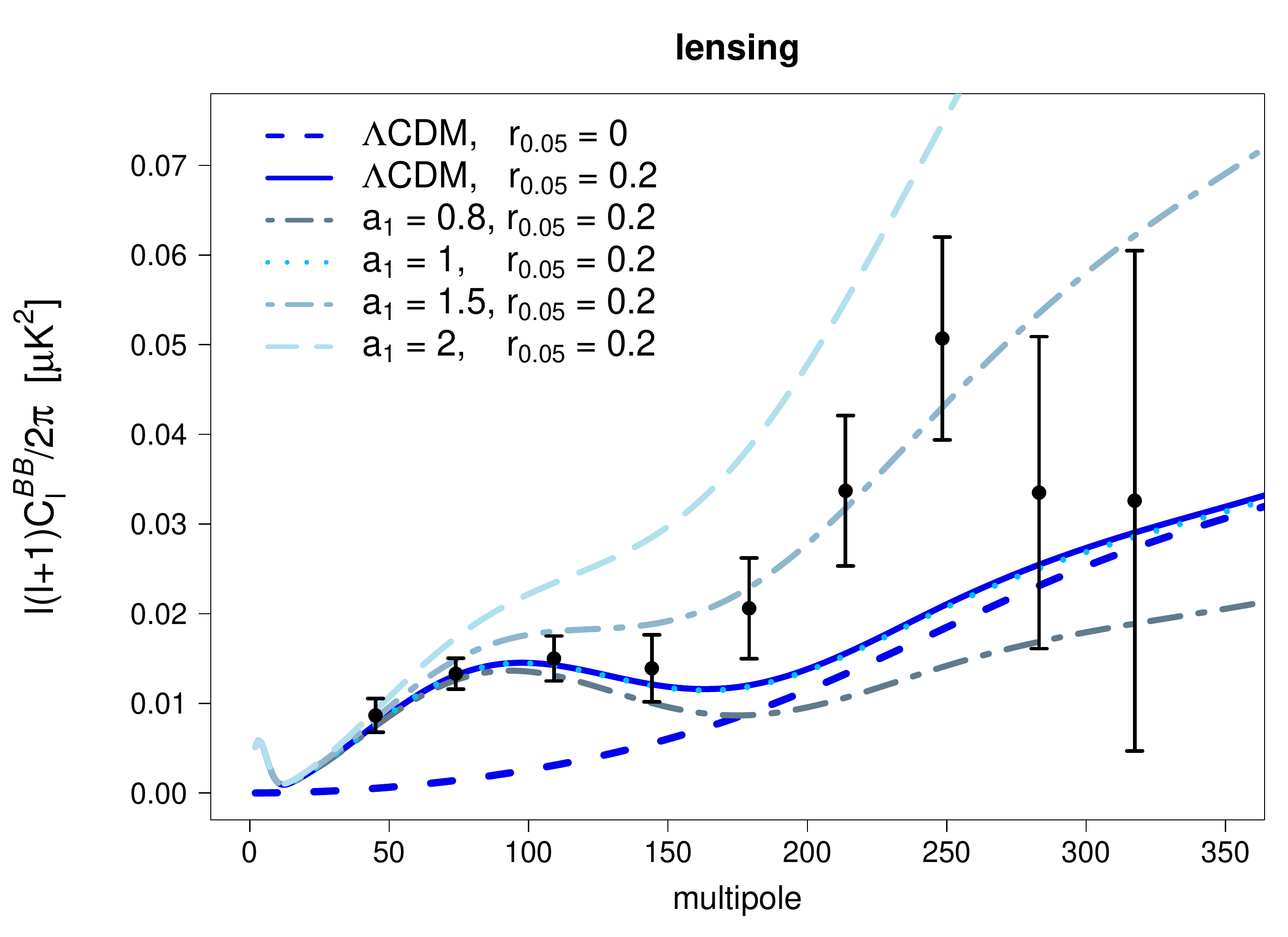}\tabularnewline
 & \tabularnewline
\includegraphics[width=85mm]{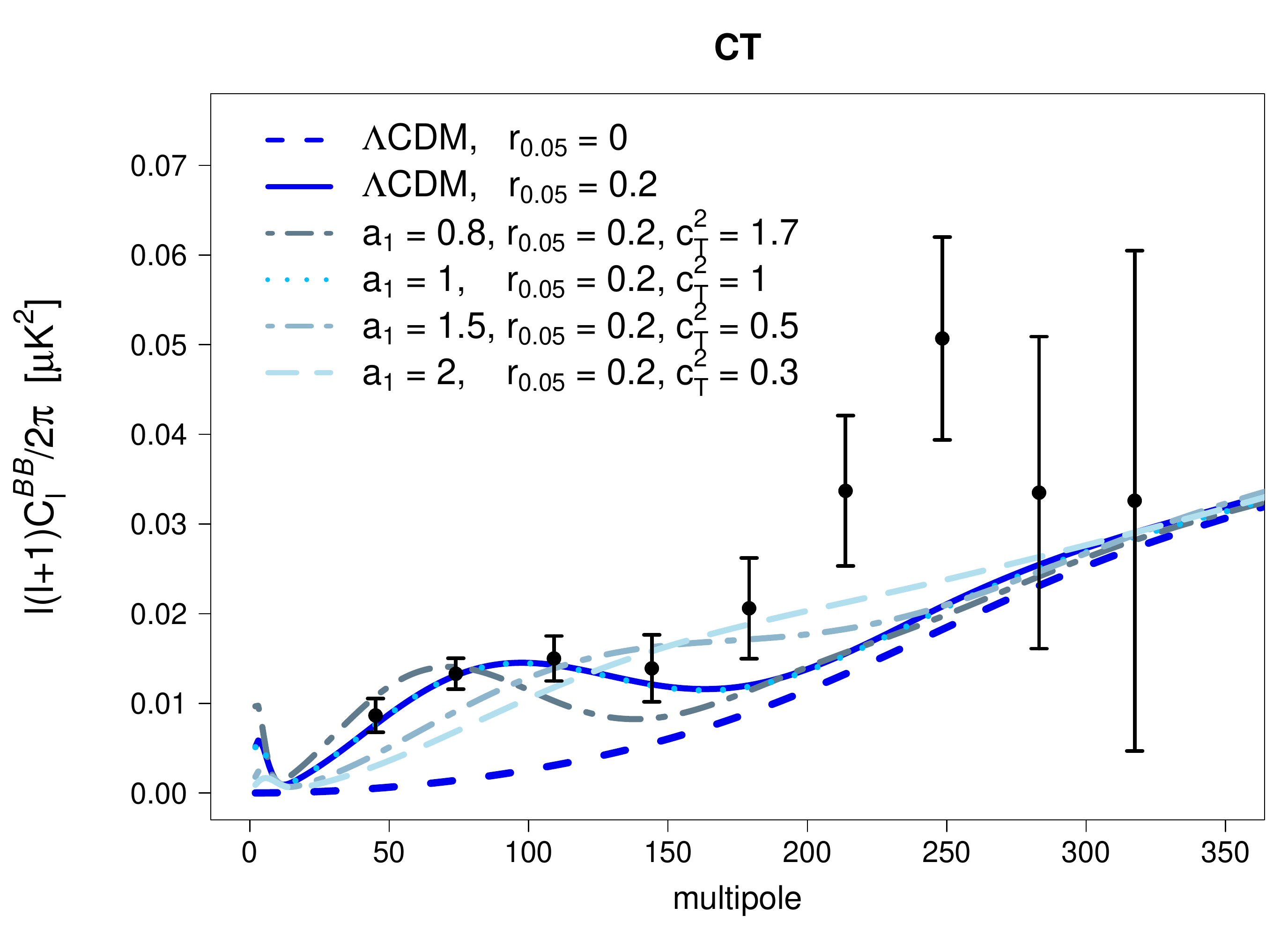}  & \includegraphics[width=85mm]{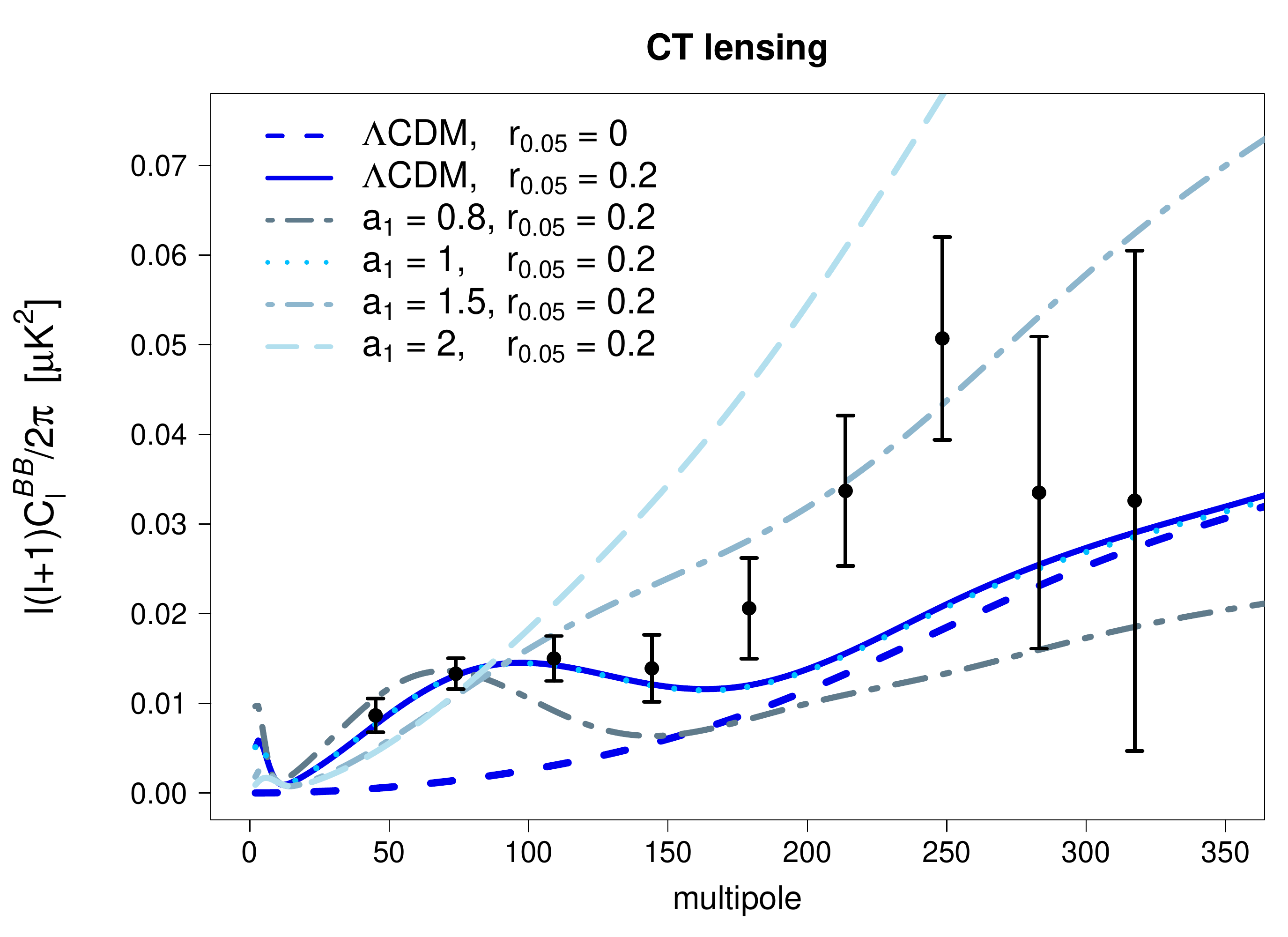}\tabularnewline
\end{tabular}\caption{BB power spectrum for MG models. In the top panels we show the effect
of a lensing correction for $r=0$ and $r=0.2$ respectively. In the
bottom left panel we plot the case in which the 'CT' correction on
the speed of gravitational waves is active (while the lensing is standard).
In the bottom right panel we activate both effects. In all cases for
simplicity we assume $a_{2}=a_{3}$ and different values of the $a_{1}$
parameter. The corresponding values of $c_T^2$ are written in the bottom left panel and are related to $a_1$ via eq.(\ref{gwct}).
 For comparison, the predictions for $\Lambda$CDM with $r=0$ (short
dashed, blue) and $r=0.2$ (solid, blue) are also shown. The black
dots are the data points from BICEP2. }

\label{fig:BB_all} 
\end{figure}

\begin{figure}[!]
\centering \includegraphics[width=0.633\textwidth]{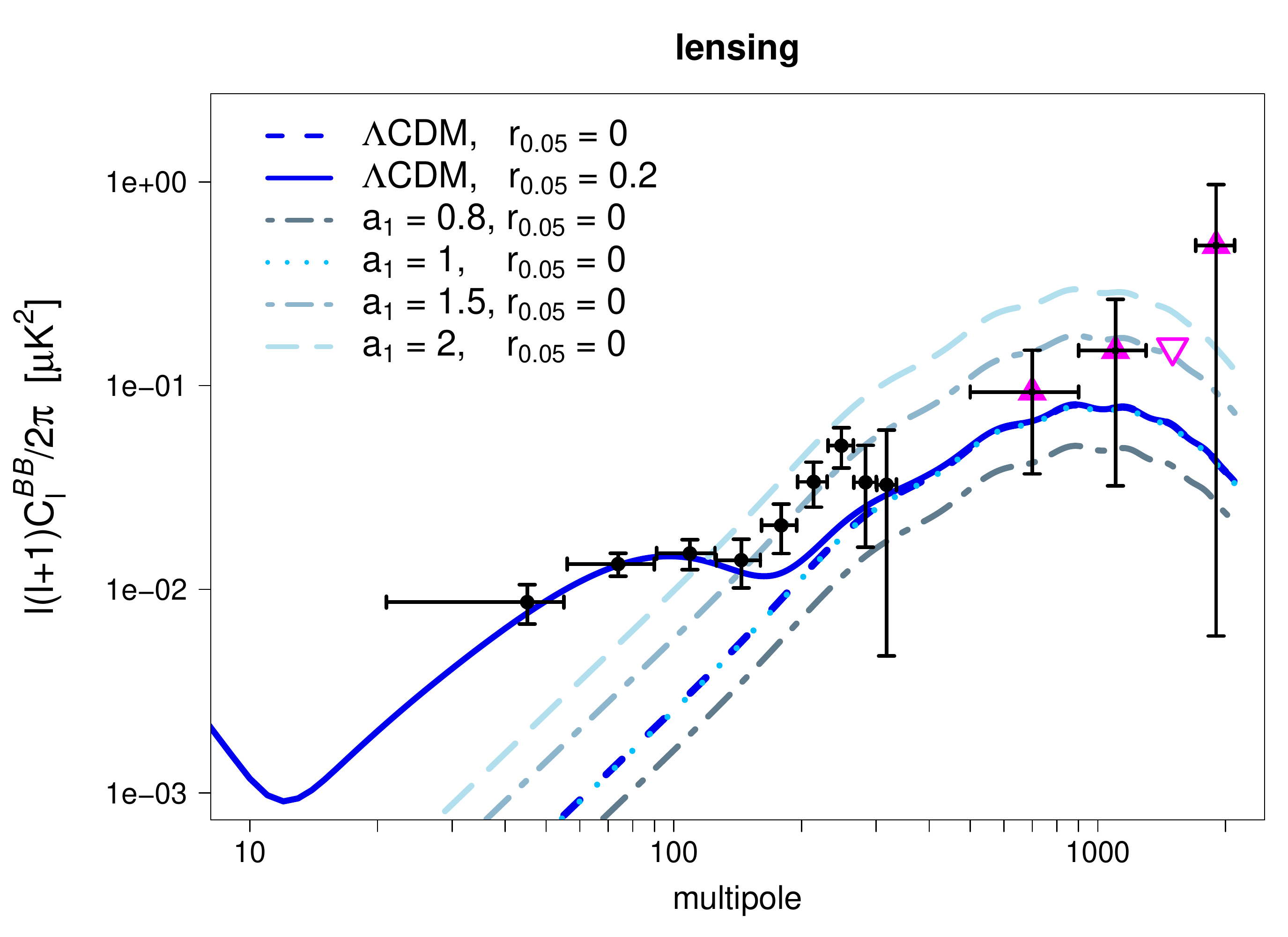}
\caption{The theoretical predictions of Fig. (\ref{fig:BB_all}), first panel,
are shown here for a larger $\ell$ range with a $\log$ scale. The
horizontal error bars associated to the BICEP2 data points correspond
to the interval $(\ell_{min},\ell_{max})$ from the data currently
available from the BICEP2 collaboration. For reference, although not
used in this analysis, we also over plot data from POLARBEAR (magenta,
triangular points). The third of these points is given as an upper
limit at 2 standard deviations.}

\label{fig:BB_largel} \includegraphics[width=0.63\textwidth]{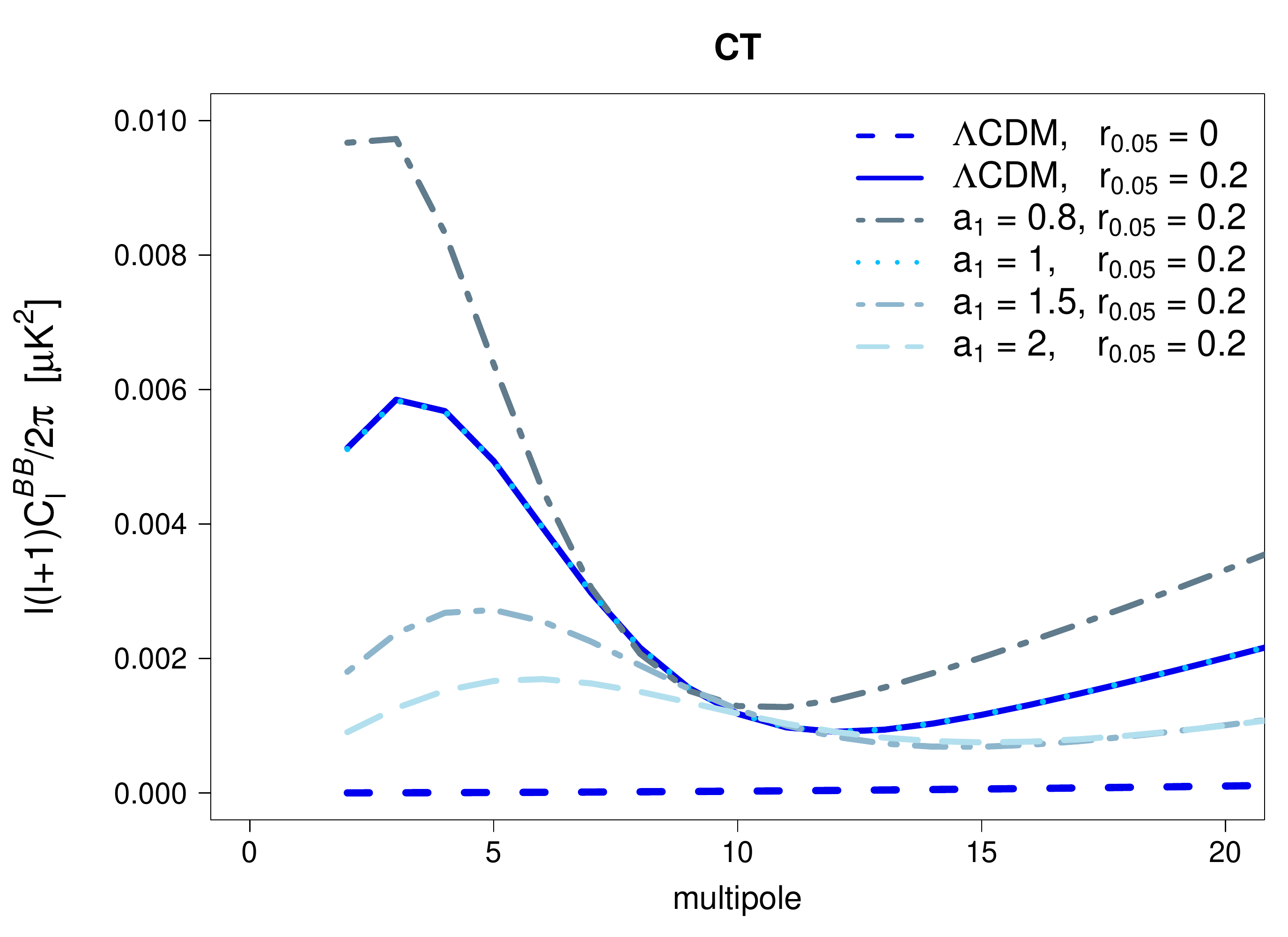}
\caption{Zoom at low multipoles of the BB spectrum of Fig.(\ref{fig:BB_all}),
bottom left panel. A CT effect, i.e. a change in the speed of gravitational
waves, is expected to modify the reionization peak in B-modes. }

\label{fig:BB} 
\end{figure}

To illustrate the effects of MG on the BB polarization we show its
spectrum 
in Figures (\ref{fig:BB_all} - 
\ref{fig:BB}) for various choices of $a_{1}$ while we always fix
$a_{2}=a_{3}$, just for illustration.

In Fig.(\ref{fig:BB_all}) we plot the BB spectrum for the three
effects (lensing, CT, CT + lensing). The top panels refer to the lensing
case for a tensor to scalar ratio $r_{0.05}=0$ and $r_{0.05}=0.2$
respectively. The $\Lambda$CDM is also shown for reference for both
values of $r_{0.05}$. As expected, increasing the value of $a_{1}$
effectively increases $\eta$ and therefore the integrand in eq.(\ref{cll}):
the lensing contribution therefore increases in amplitude and extends
to smaller multipoles than expected for the corresponding $\Lambda$CDM.
The position of the primordial peak is however not affected (although
its amplitude also receives a contribution from MG). The bottom left
panel shows the BB spectra when only the CT effect is active: in this
case, the speed of gravitational waves changes with the inverse of
$a_{1}$ as described in eq. (\ref{gwct}): a value of $a_{1}$ smaller
(larger) than one increases (decreases) the speed of gravitational
waves with respect to $\Lambda$CDM and shifts the expected position
of the primordial peak to smaller (larger) multipoles. When both effects are active, as in the bottom right
panel, both the lensing amplitude and the shift in peak position can
occur. In all panels we also overplot for reference  the recently released
data points of BICEP2.%
\footnote{http://bicepkeck.org/; B2\_3yr\_bandpowers\_20140314.txt
} We note that the corresponding TT and EE spectra are very little
affected by these changes,  so that in our model the main contribution of
 MG is on B-modes.
In Figure (\ref{fig:BB_largel}) we replot the lensing case, as in
Fig. (\ref{fig:BB_all}) top left panel, for $r_{0.05}=0$ and a wider
range in scales, up to $\ell=2000$, to show how the lensing predictions
for MG compare with the available data from POLARBEAR,%
\footnote{http://lambda.gsfc.nasa.gov/product/suborbit/polarbear\_prod\_table.cfm
} although we do not use them in our analysis. Notice that if we took
BICEP2 data at face value, while the ``bump'' around $\ell\sim80$
is best rendered by a non-zero tensor-to-scalar ratio ($r=0.2$) in
$\Lambda$CDM, the location of the upper points is qualitatively in
good agreement with a modification of gravity represented by $a_1 \approx1.5$.

Finally, in Figure (\ref{fig:BB}), we zoom in the low-$\ell$ region
in order to emphasize the effect of a change in $c_{T}$ on the reionization
peak. Interestingly, future measurements of the reionization peak
by experiments like those in Refs.\cite{2012SPIE.8442E..19H,2009arXiv0906.1188B,2011JCAP...07..025K,2013arXiv1306.2259P,2011arXiv1102.2181T} could be used to discriminate MG theories. For example, the  satellite mission LiteBIRD  \cite{2013arXiv1311.2847M} would have a sensitivity to characterise the tensor to scalar ratio r with an uncertainty of $\delta r \sim 0.001$. At these scales the lensing contribution is negligible
and the only modification  comes from the correction in the
speed of gravitational waves.

 We then proceed with Monte Carlo simulations using available data to test the parameters of
our implementation of MG.  We use WMAP9 data \cite{wmap9},
ACT \cite{das_etal_2013} + SPT \cite{reichardt_etal_2012} and the data from BICEP2 \cite{Ade:2014xna} on B-modes
polarization. We enforce the inflationary consistency relation $n_t\simeq -r/8$ for the
tensor spectral index $n_{t}$. We perform different runs, including
the three cases illustrated before: lensing modification, CT, CT +
lensing modifications. Results are illustrated in Table \ref{Tab:values}.
As the values for $a_{2}$ and $a_{3}$ can in principle span several
orders of magnitude, we use for convenience the logarithm of these
quantities. The parameters $a_{2}$ and $a_{3}$ play no role in CT and
are essentially unconstrained also including the lensing MG effect. Provided that the
foreground contributions are under control, B-modes polarization will
be however a very good probe to test the $a_{1}$ parameter, i.e. the
anisotropic stress and the speed of gravitational waves.

\begin{table}[htdp]
\begin{centering}
\begin{tabular}{|c|c|c|c|}
\hline 
\,\,\,\,\,\,\,\,\,\,\,\,\,\,\,\,\,\,\,\,\,\,\,\,\,\,\,\,\,\,\,\,\,\,\,\,\,\,\,\,\,\,\,\,\,
Datasets: WMAP9 + HighL + BICEP2\,\,\,\,\,\,\,\,\,\,\,\,\,\,\,\,\,\,\,\,\,\,\,\,\,\,\,\,\,\,\,\,\,\,\,\,\,\,\,\,\,\,\,\,   \tabularnewline
\hline 
\end{tabular}\\
\begin{tabular}{|c|c|c|c|c|c|}
\hline 
Parameter  & $\Lambda$CDM  & MG (lensing)  & MG (CT)  & MG (CT + lensing)   \tabularnewline
\hline 
$a_{1}$  & -  & 1.30 $\pm$ 0.16  & $2.89\pm1.21$  & $1.28\pm0.15$   \tabularnewline
\hline 
$\log_{10}\, a_{2}$  & -  & < 0.30  & -  & $<-1.12$   \tabularnewline
\hline 
$\log_{10}\, a_{3}$  & -  & < 0.37  & -  & $<-0.11$   \tabularnewline
\hline 
$r_{0.05}$  & 0.21 $\pm$ 0.05  & 0.19 $\pm$ 0.05  & $0.35\pm0.11$  & $0.21\pm0.05$  \tabularnewline
\hline 
$r_{0.002}$  & 0.23 $\pm$ 0.06  & 0.21 $\pm$ 0.06  & $0.43\pm0.17$  & $0.23\pm0.07$   \tabularnewline
\hline 
$H_{0}$  & 72.0 $\pm$ 2.2  & 73.7 $\pm$ 2.4  & $73.0\pm2.4$  & $73.9\pm2.4$   \tabularnewline
\hline 
$n_{s}$  & 0.998 $\pm$ 0.013  & 1.000 $\pm$ 0.014  & $1.012\pm0.016$  & $1.006\pm0.015$  \tabularnewline
\hline 
\hline 
$-\log\mathcal{L}$  & 4146  & 4142  & 4145  & 4143   \tabularnewline
\hline 
\end{tabular}
\par\end{centering}

\caption{Mean values $\pm$ standard deviation for a selection of parameters.
Both columns refer to the combination of datasets WMAP + HighL + BICEP2.
For $\log(a_{2})$ and $\log(a_{3})$ parameters we write the 95$\%$
upper limit; these parameters are essentially not constrained. }

\label{Tab:values} 
\end{table}

\begin{figure}
\begin{centering}
\includegraphics[width=0.8\textwidth]{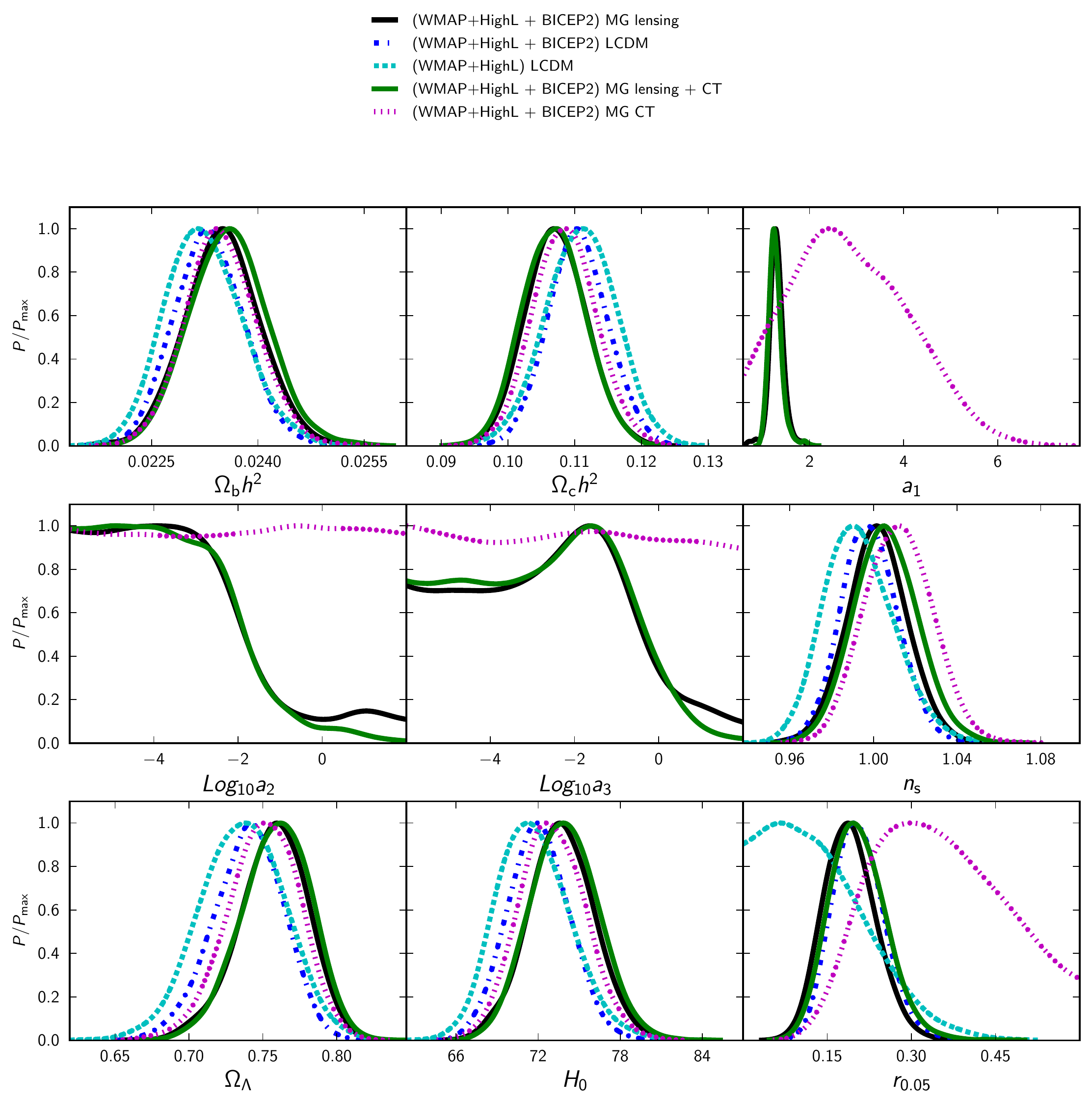}
\caption{One-dimensional posterior contours for a selection of cosmological
parameters. We compare the three cases in which: lensing is modified
(solid, black line), CT is modified (dotted magenta line), CT and
lensing are both modified (light solid, green line). We over plot
also the case of $\Lambda$CDM for comparison (dot dashed, blue line).
}
\label{fig:1d_abc} 
\par\end{centering}
\centering{} 
\end{figure}

In Fig.(\ref{fig:1d_abc}) we show 1D posterior contours for a selection
of cosmological parameters and different runs. Here it becomes clearer
that $a_{1}$ is mainly constrained by the lensing contribution, while
CT gives much larger uncertainty. This can be seen intuitively from
Fig.(\ref{fig:BB_all}): CT influences both the amplitude and the
position of the BB spectrum and is able to fit the data for a larger
range in $a_{1}$.

\begin{figure}
\begin{centering}
\includegraphics[width=0.8\textwidth]{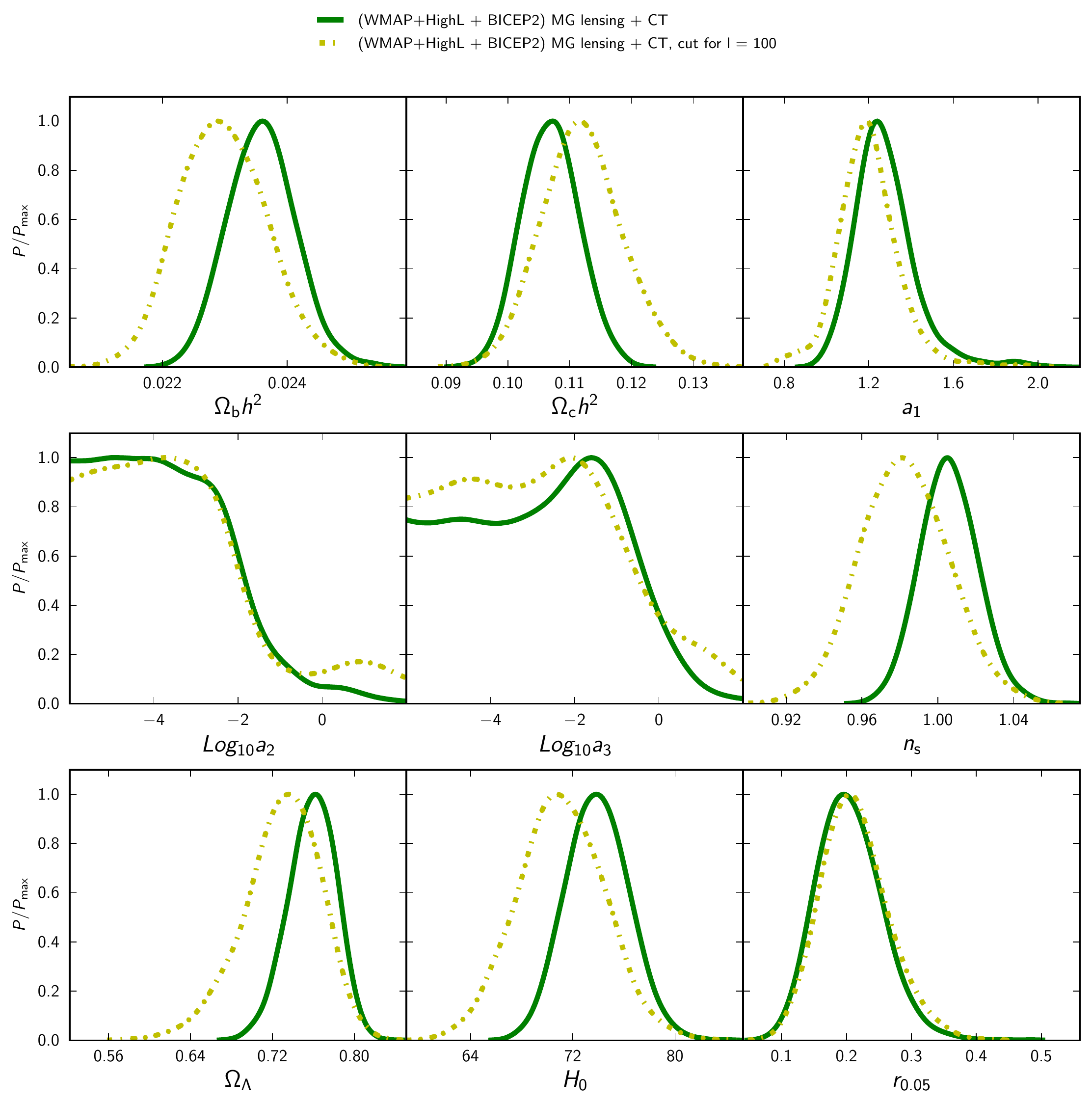}
\caption{One-dimensional posterior contours for a selection of cosmological
parameters. For the case in which both CT and lensing are both modified,
we compare the constraints obtained using all multipoles (solid, green
line), to the case in which we cut all multipoles $\ell<100$ in TT and TE (dash-dotted, light green).}
\label{fig:1d_abc_cut} 
\par\end{centering}
\centering{}
\end{figure}

In order to check the validity of the QS limit, we test
the effect of the incorrectly modelled low-$\ell$ part of the TT
spectrum by redoing the MCMC simulation cutting the first 100 multipoles
in the TT and TE spectra. A comparison of the run with and without cut
is shown in Fig.(\ref{fig:1d_abc_cut}). As we see, while parameters
like $n_{s}$ and $H_{0}$ are affected by the cut, constraints on
$a_{1}$  do not depend on the cut; i.e. they
do not depend much on the low $\ell$ multipoles in the temperature
spectra. This is reassuring as it shows that the QS limit
and its simplest numerical implementation may be sufficient to test
MG theories.

\begin{figure}[htp]
\centering %
\begin{tabular}{ccc}
\includegraphics[width=55mm]{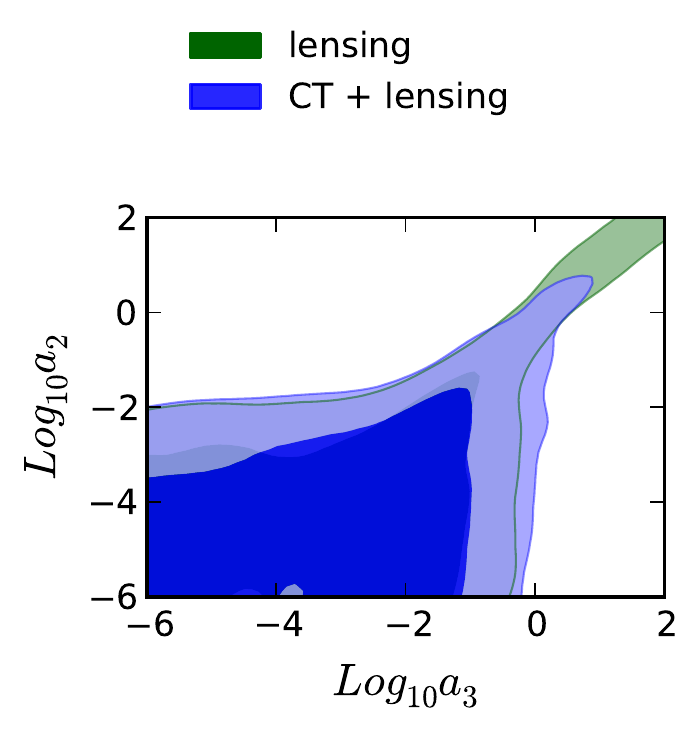}  & 
\includegraphics[width=55mm]{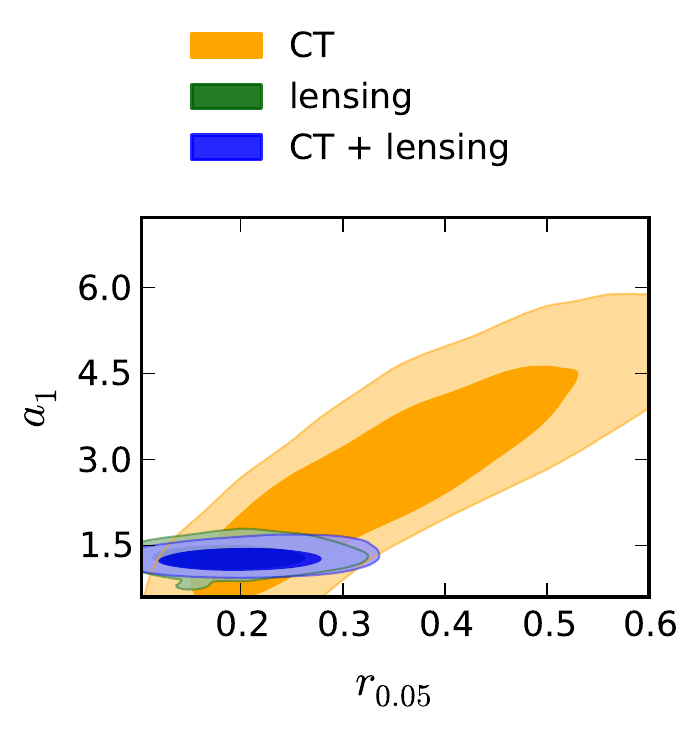}  & 
\includegraphics[width=55mm]{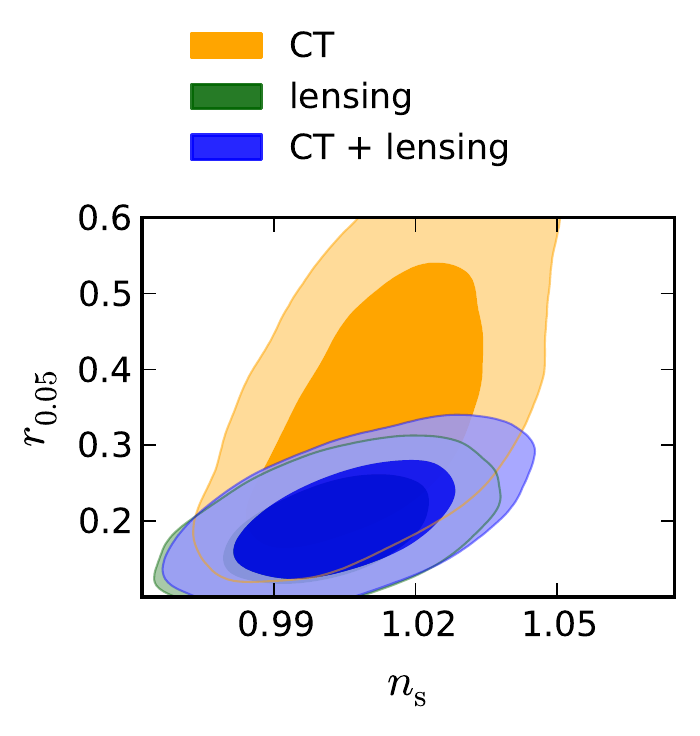}\tabularnewline
\end{tabular}\caption{2D posterior contours for a selection of cosmological parameters and
the three cases in which lensing only is modified (green contours),
CT only is modified (orange/light contours), CT and lensing are both
modified (blue/darker contours). The case including CT
only does not depend on $a_{2}$ and $a_{3}$ parameters. As before,
we consider the data combination WMAP + HighL+ BICEP2.}

\label{Fig:2d} 
\end{figure}

In Fig.(\ref{Fig:2d}) we show the comparison of the 2D posterior
contours for the three effects. In addition to the
considerations already done above, we notice  here that $a_{2}$
and $a_{3}$ are degenerate and tend to align along the direction
$a_{2}=a_{3}$. This particular direction removes any scale dependence in
the expression for $\eta$ in (\ref{def_sigma}).

\begin{figure}
\begin{centering}
\includegraphics[width=0.6\textwidth]{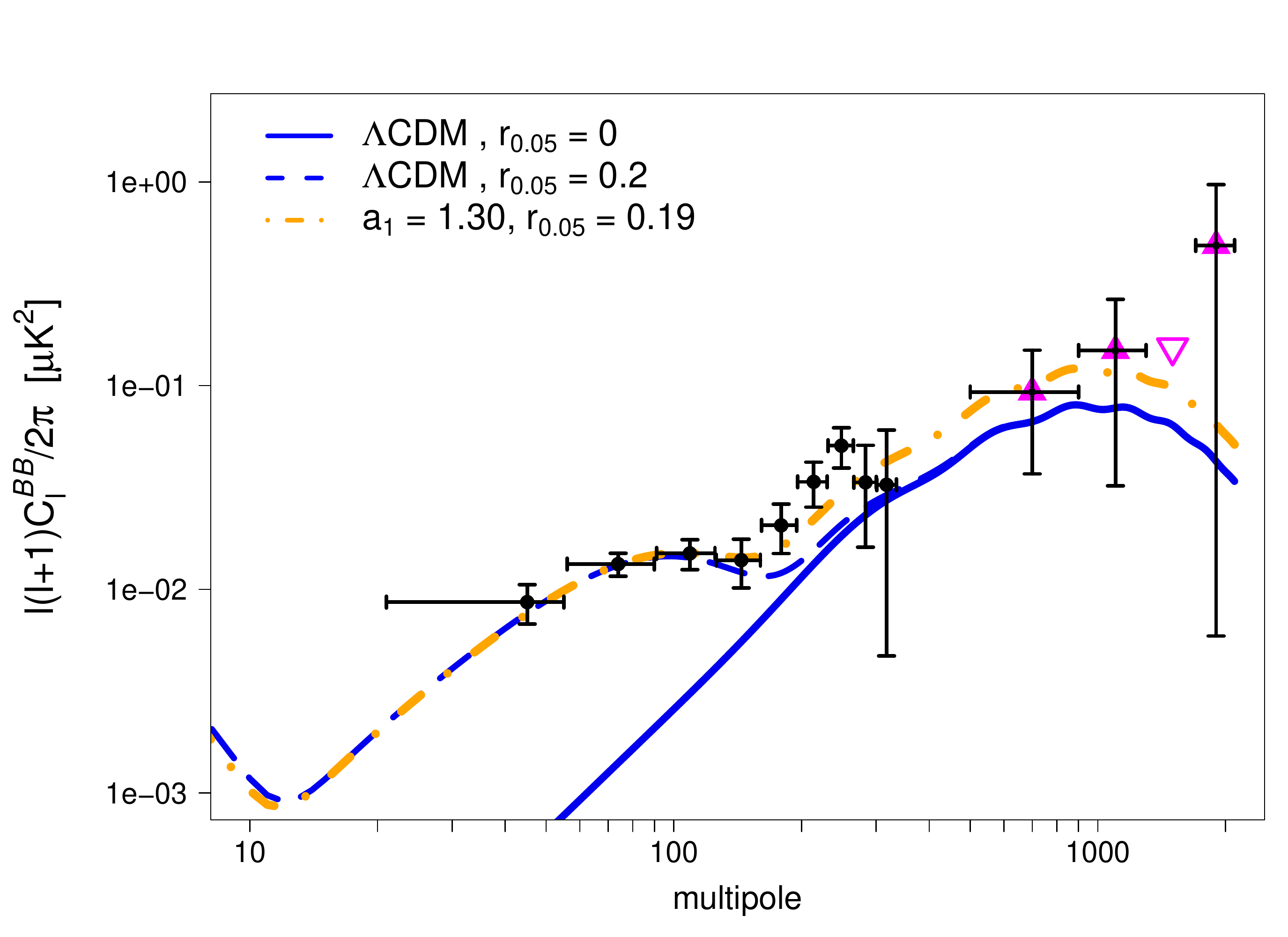}
\caption{Best fits from Table 1. MG with $r\neq0$ is shown (dash-dotted, orange)
together with $\Lambda$CDM assuming also $r\neq0$ (dashed, blue). 
For comparison we also show $\Lambda$CDM with $r=0$ (blue). As in
Figure 2, we show the data points of POLARBEAR (magenta triangles)
in addition to those of BICEP2. The error in the abscissa associated
to the BICEP2 data points corresponds to the interval $(\ell_{min},\ell_{max})$
from the release by the BICEP2 team. The third point from POLARBEAR
is plotted as an upper limit at 2 standard deviations. We recall that
POLARBEAR data are not used in the analysis and are only shown here
for reference.}
\label{fig:BB_largel_bestMG} 
\par\end{centering}
\begin{centering}
\par\end{centering}
\centering{} 
\end{figure}

In Fig.(\ref{fig:BB_largel_bestMG}) we redo Fig.(\ref{fig:BB_largel})
for $a_1=1.30$, corresponding to the mean (and best fit) of modified gravity, for the lensing case shown in the plot.

\begin{figure}[htp]
\centering %
\begin{tabular}{ccc}
\includegraphics[width=95mm]{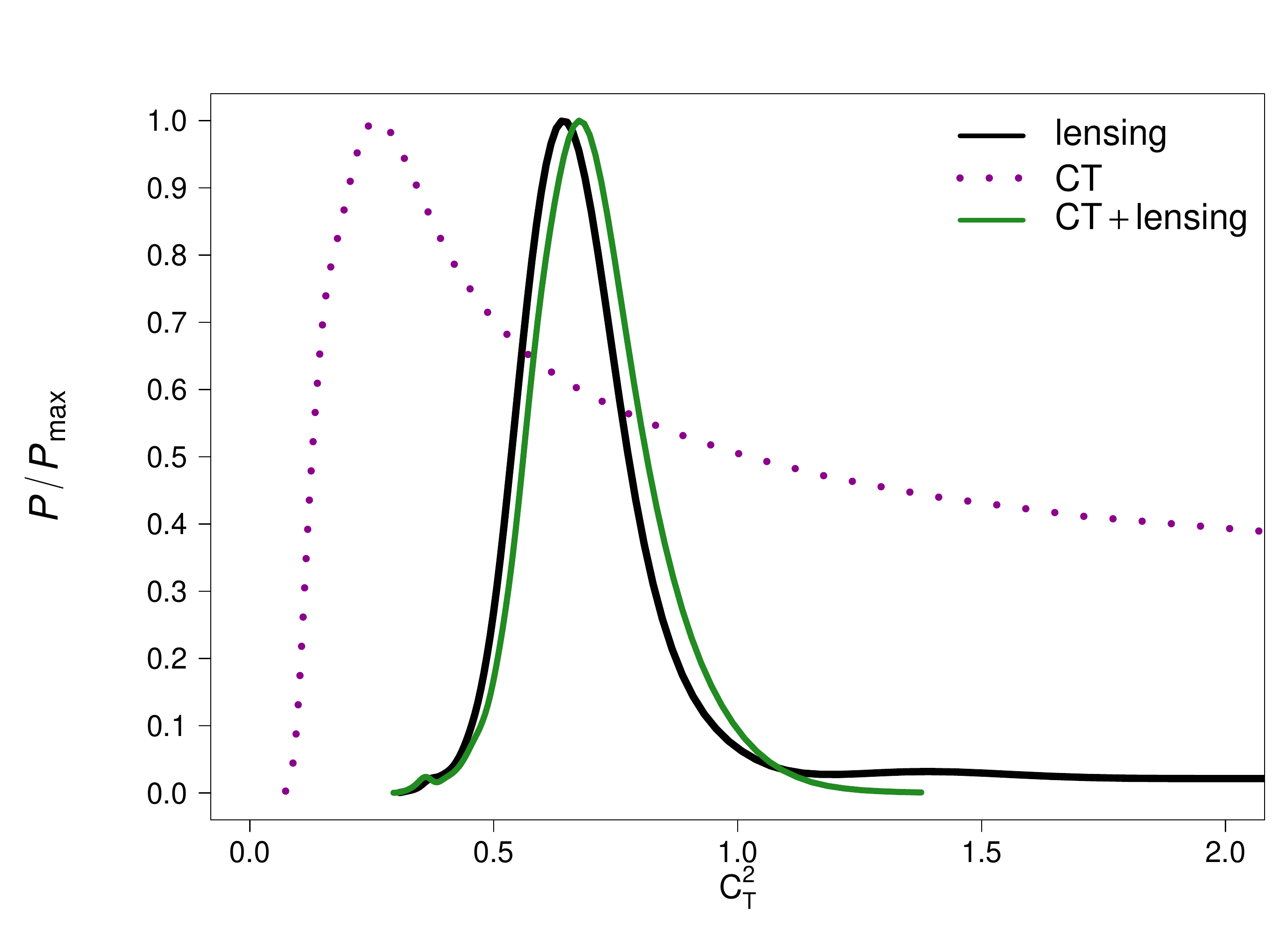}  &  & \tabularnewline
\end{tabular}\caption{Speed of gravitational waves, as obtained remapping the constraints
on $a_{1}$ for the three different effects considered in this paper:
lensing (dark solid, black line), CT (dotted magenta), CT + lensing
(lighter solid, green). $c_{T}^2=1$ corresponds to the standard value.}

\label{Fig:ct_remapped} 
\end{figure}

Finally, we remap the constraints obtained for the various runs on
$a_{1}$ into $c_{T}$, the speed of gravitational waves. The resulting
1D contours are shown in Fig.(\ref{Fig:ct_remapped}). We find that
$c_{T}=0.8\pm 0.07$ from CT+lensing. Using CT alone the constraint is much weaker: $c_T \gtrsim 0.4$ at 2$\sigma$.
The reason for this behavior can be understood by looking at Fig. \ref{Fig:2d}. In the central panel one can see
that $a_1$ (or equivalently $c_T$) is quite degenerate with $r_{0.05}$ if lensing is not taken into account. This is because a shift of the tensor peak towards higher $\ell$s can be partially 
compensated by an increase of $r_{0.05}$. In other words, $a_1$ (or $c_T$) could be determined by 
%are measured only through 
the tensor peak position (see the bottom-left panel of Fig. 1) which is not firmly established by the current data. However,
%In contrast, 
$c_T$ changes the lensing amplitude in a significant way (see Fig. \ref{fig:BB_all}, top-right panel) and is therefore  well measured by lensing alone.

\section{Conclusions}

\label{conc}

The polarized light from the last scattering surface carries important
information in addition to the temperature anisotropy. Not only it
helps constraining the cosmological parameters but it also allows
to separate the effects of primordial gravitational waves from scalar
perturbations, both predicted by inflationary models. Current experiments
\cite{Ade:2013gez,Ade:2014xna, ActPol2014} are already providing
results which will soon be cross tested by Planck. Future observations \cite{2012SPIE.8442E..19H,2009arXiv0906.1188B,2011JCAP...07..025K,2013arXiv1306.2259P,2011arXiv1102.2181T}  
will  keep improving our knowledge
of CMB polarization. B-modes, once foregrounds are accounted for, are particularly important in this context
since they contain both the imprint of primordial gravitational waves 
and the one from gravitational lensing induced by large-scale structure. In
a sense, B-modes are the ultimate test of gravity at cosmological
scales since they probe  two genuinely general relativistic
effects.

Modifications of gravity have been mainly proposed to describe the late time acceleration of the universe (see e.g. \cite{Amendola2010a} for a review) but they have also been studied as a possibility for driving primordial inflation (e.g. \cite{Kobayashi:2011nu, Gao:2011qe, DeFelice:2011uc, Tsujikawa:2013ila}). In any case, it is conceivable that some extra degrees of freedom affect gravity at early times, see for instance \cite{Konnig:2014xva}. In this paper we investigated how B-modes can be employed to test gravity at early times and at cosmological scales.  Both lensing and gravitational
wave propagation depend on the features of gravity and one can use them to constrain its properties. It is remarkable that both effects
depend on the amount of anisotropic stress
$\eta$, which in general differs from the standard value of unity in
modified gravity. To simplify our study, we single out the effects
that depend on $\eta$ alone by selecting models in which the
background and the matter perturbation growth are exactly standard.
Moreover, we work in the quasi-static limit, in which all
the modified gravity effects can be embodied in just two arbitrary
functions of time and space. In a vast class of models (the Horndeski Lagrangian \cite{Horndeski:1974} and 
in bimetric gravity \cite{Hassan:2011zd}), these two functions
take a particularly simple form, given by eq.(\ref{limo1}).

We show by a Monte Carlo analysis with real data that it is
indeed possible to constrain the anisotropic stress and the gravitational
wave speed with B-modes. Although the particular values we obtain
here are to be taken with great caution because the B-mode available data are
still under close scrutiny,  future data  has great potential for providing tight constraints on MG.\footnote{Shortly after our work was made public, a related study on the speed of gravitational waves also appeared, see \cite{Raveri:2014eea}.}

 Several
of the assumptions we adopted for simplicity in this work can be lifted relatively
easily: one can for instance remove the assumption of constant
MG parameters and work with the full equations rather than with their quasi
static limit. This will be addressed in future works. 

\begin{acknowledgments}
We acknowledge support from the TRR33 - The Dark Universe - DFG Grant.
We thank Emilio Bellini, Diego Blas, Alicia Bueno-Belloso, Martin Kunz,  Ippocratis Saltas, 
Ignacy Sawicki, and Licia Verde for fruitful discussions. We also thank Francesco Montanari for useful comments on a draft version of this work. 
\end{acknowledgments}

\appendix
%dummy comment inserted by tex2lyx to ensure that this paragraph is not empty
%dummy comment inserted by tex2lyx to ensure that this paragraph is not empty
%dummy comment inserted by tex2lyx to ensure that this paragraph is not empty
%dummy comment inserted by tex2lyx to ensure that this paragraph is not empty

\section{Anisotropic stress without modified growth}

\label{appe} One may wonder whether taking $\eta\neq1$ and $Y=1$
at the same time is consistent since it can be expected that a general
modification of gravity would induce a change in both Poisson's equation
and the relation between the metric potentials, $\Psi$ and $\Phi$.
Although this is indeed the general case, it is possible to find situations
in the QS limit that produce an anisotropic stress component, but
do not modify Newton's constant. We will illustrate this with an example. 

The action 
\begin{equation}
S=\int\text{d}^{4}x\sqrt{-g}\left(\tfrac{1}{2}R+\mathcal{L}_{x}+\mathcal{L}_{\text{m}}\right)\,,\label{dynamics}
\end{equation}
describes the behaviour of matter (with Lagrangian density $\mathcal{L}_{\text{m}}$)
in a modification of GR given by $\mathcal{L}_{\text{x}}$. In the
case of Hordenski MG theories, the modification of GR is given by
a scalar field $\phi$ with $\mathcal{L}_{\text{x}}=\sum_{i=2}^{5}\mathcal{L}_{i}$,
where 
\begin{align}
\mathcal{L}_{2}= & K(\phi,X)\,,\\
\mathcal{L}_{3}= & -G_{3}(\phi,X)\Box\phi\,,\\
\mathcal{L}_{4}= & G_{4}(\phi,X)R+G_{4,X}\left[\left(\Box\phi\right)^{2}-\left(\nabla_{\mu}\nabla_{\nu}\phi\right)\left(\nabla^{\mu}\nabla^{\nu}\phi\right)\right]\,,\\
\mathcal{L}_{5}= & G_{5}(\phi,X)G_{\mu\nu}\nabla^{\mu}\nabla^{\nu}\phi-\frac{G_{5,X}}{6}\Bigl[\left(\Box\phi\right)^{3}-3\left(\Box\phi\right)\left(\nabla_{\mu}\nabla_{\nu}\phi\right)\left(\nabla^{\mu}\nabla^{\nu}\phi\right)+2\left(\nabla^{\mu}\nabla_{\alpha}\phi\right)\left(\nabla^{\alpha}\nabla_{\beta}\phi\right)\left(\nabla^{\beta}\nabla_{\mu}\phi\right)\Bigr]\,.
\end{align}
The functions $K$ and $G_{i}$ are in principle arbitrary and $X=-\partial_{\mu}\phi\,\partial^{\mu}\phi/2$
is the standard kinetic term. 

The QS limit of the equations of motion derived from this action are
characterized by the functions $h_{i}$ appearing in eq.(\ref{limo1}). They can
be expressed as 
\begin{align}
h_{1} & \equiv\frac{w_{4}}{w_{1}^{2}}=\frac{c_{\text{T}}^{2}}{w_{1}}\,,\qquad h_{2}\equiv\frac{w_{1}}{w_{4}}=c_{\text{T}}^{-2}\,,\label{eq:h1h2}\\
h_{3} & \equiv\frac{H^{2}}{2XM^{2}}\frac{2w_{1}^{2}w_{2}H-w_{2}^{2}w_{4}+4w_{1}w_{2}\dot{w}_{1}-2w_{1}^{2}(\dot{w}_{2}+\rho_{\text{m}})}{2w_{1}^{2}}\,,\nonumber \\
h_{4} & \equiv\frac{H^{2}}{2XM^{2}}\frac{2w_{1}^{2}H^{2}-w_{2}w_{4}H+2w_{1}\dot{w}_{1}H+w_{2}\dot{w_{1}}-w_{1}(\dot{w}_{2}+\rho_{\text{m}})}{w_{1}}\,,\nonumber \\
h_{5} & \equiv\frac{H^{2}}{2XM^{2}}\frac{2w_{1}^{2}H^{2}-w_{2}w_{4}H+4w_{1}\dot{w}_{1}H+2\dot{w_{1}}^{2}-w_{4}(\dot{w}_{2}+\rho_{\text{m}})}{w_{4}}\,,\nonumber 
\end{align}
being 
\begin{align}
w_{1}\equiv & 1+2\left(G_{4}-2XG_{4,X}+XG_{5,\phi}-\dot{\phi}XHG_{5,X}\right)\,,\label{eq:ws}\\
w_{2}\equiv & -2\dot{\phi}\left(XG_{3,X}-G_{4,\phi}-2XG_{4,\phi X}\right)+\nonumber \\
 & +2H\left(w_{1}-4X\left(G_{4,X}+2XG_{4,XX}-G_{5,\phi}-XG_{5,\phi X}\right)\right)-\nonumber \\
 & -2\dot{\phi}XH^{2}\left(3G_{5,X}+2XG_{5,XX}\right)\,,\nonumber \\
w_{3}\equiv & 3X\left(K_{,X}+2XK_{,XX}-2G_{3,\phi}-2XG_{3,\phi X}\right)+18\dot{\phi}XH\left(2G_{3,X}+XG_{3,XX}\right)-\nonumber \\
 & -18\dot{\phi}H\left(G_{4,\phi}+5XG_{4,\phi X}+2X^{2}G_{4,\phi XX}\right)-\nonumber \\
 & -18H^{2}\left(1+G_{4}-7XG_{4,X}-16X^{2}G_{4,XX}-4X^{3}G_{4,XXX}\right)-\nonumber \\
 & -18XH^{2}\Bigl(6G_{5,\phi}+9XG_{5,\phi X}+2X^{2}G_{5,\phi XX}\Bigr)+\nonumber \\
 & +6\dot{\phi}XH^{3}\Bigl(15G_{5,X}+13XG_{5,XX}+2X^{2}G_{5,XXX}\Bigr)\,,\nonumber \\
w_{4}\equiv & 1+2\left(G_{4}-XG_{5,\phi}-XG_{5,X}\ddot{\phi}\right)\,.\nonumber 
\end{align}
and 
\begin{align}
M^{2}{\dot{\phi}} & =3H\left(P_{x,\phi}+\rho_{x,\phi}\right)+\dot{\rho}_{x,\phi}\\
\rho_{x} & =3H^{2}(1-w_{1})+2XK_{,X}-K-2XG_{3,\phi}+6\dot{\phi}H\left(XG_{3,X}-G_{4,\phi}-2XG_{4,\phi X}\right)+\nonumber \\
 & \,\quad+12H^{2} X \left(G_{4,X}+2XG_{4,XX}-G_{5,\phi}-XG_{5,\phi X}\right)+4\dot{\phi}XH^{3}\left(G_{5,X}+XG_{5,XX}\right)\,,\nonumber \\
P_{x} & =-\left(3H^{2}+2\dot{H}\right)(1-w_{1})+K-2XG_{3,\phi}+4XG_{4,\phi\phi}+2\dot{\phi}Hw_{1,\phi}-4X^{2}H^{2}G_{5,\phi X}\\
 & \,\quad+2\dot{\phi}XH^{3}G_{5,X}+\ddot{\phi}\left(w_{2}-2Hw_{1}\right)/{\dot{\phi}}\,.\nonumber 
\end{align}

We see from eq. (\ref{Y_def}) that in order to get $\eta\neq1$ and
$Y=1$ in the QS limit we must impose 
\begin{align}
h_{3} & =h_{5}\label{c1}\\
h_{1} & =1\,.\label{c2}
\end{align}
The condition (\ref{c2}) enforces 
\begin{align}
w_{4}=w_{1}^{2}\,.\label{eq:condh1}
\end{align}
Then, the condition (\ref{c1}) becomes a relation between the Hubble
parameter $H$ and the functions $w_{1}$ and $w_{2}$: 
\begin{align}
\dot{w}_{1}=w_{1}\left(\frac{w_{2}}{2}-H\right)\,.\label{cond}
\end{align}

If we impose
\begin{equation}
G_{3,X}=0\,\quad G_{4,\phi}=0\,\quad G_{5,X}=0\,,\label{eq:conds}
\end{equation}
the equations $w_{2}=2H$ and $w_{4}=w_{1}^{2}$ are equivalent to 
\begin{align}
G_{4}+X(3G_{5,\phi}-4G_{4,X}-4XG_{4,XX}) & =0\label{a1}\\
2(G_{4}-XG_{4,X})^{2}+X^{2}(9+8G_{4}-8XG_{4,X})G_{4,XX}+8X^{4}G_{4,XX}^{2} & =0\,.\label{eq:newc1}
\end{align}
Under these conditions, the function $w_{1}$ takes the form 
\begin{align}
w_{1}=1+\frac{4}{3}\left(G_{4}-XG_{4,X}+2X^{2}G_{4,XX}\right)\,.\label{a2}
\end{align}
Then, if $G_{5}\propto\phi$ and the field evolves keeping $X$ constant,
the equations (\ref{a1}) and (\ref{eq:newc1}) become algebraic constraints
and $w_{1}$ can be different from 1 (as
required to have a non-trivial $\eta$). With this choice, $Y=1$ and
$\eta=1/w_{1}$ and constant. 

Assuming that the matter component has zero pressure and $\rho_{m}$
energy density, the equations of motion in the background (the Friedmann
equations) are in this case 
\begin{align} \label{fr1}
\rho_{m}-K+2XK_{,X}-2XG_{3,\phi}=3H^{2}\,\\ \label{fr2}
\eta\left(-K+2XG_{3,\phi}\right)=3H^{2}+2\dot{H}\,.
\end{align}
Notice that if only matter is present, we recover  the standard equations $\rho_{m}=3H^{2}$ and $3H^{2}+2\dot{H}=0$. 

The equations (\ref{fr1}) and (\ref{fr2}) can be combined into 
\begin{align} \label{fr3}
\frac{d H^2}{d \ln a}+3(1+\eta)H^2=\eta\left(\rho_m-2K+2X K_{,X}\right)\,,
\end{align}
which allows to get $H$ for a given $\rho_m$ and $K(\phi,X)$. In particular, a $\Lambda$CDM evolution  for the background can be obtained provided that the matter density today is 
\begin{align} \label{omegaM}
\Omega_m^0=1-\frac{2\eta(XK_{,X}-K)}{3(1+\eta)}\,,
\end{align}
and that
\begin{equation} \label{oM2}
2(1+\eta)XG_{3,\phi}=2K_{,X}X+K(\eta-1)\,
\end{equation}
Since $\eta$ is constant, these conditions can be achieved if $G_3$ is linear in $\phi$, $K$ depends only on $X$ and the field evolves keeping $X$ constant. 
We have to check also the Klein-Gordon equation for $\phi$ (see \cite{DeFelice:2011bh}), 
which under these conditions takes the following form:
\begin{equation}
\frac{d}{d \ln a}\left(a^{3}J\right)=0, \label{eq:kg}
\end{equation}
where 
\begin{align}
J=\sqrt{2X}\left(K_{,X}-2G_{3,\phi}+6H^{2}\left(G_{4,X}+2XG_{4,XX}-G_{5,\phi}\right)\right)\,.
\end{align}
We obtain that (\ref{eq:kg}) is satisfied if $J\propto a^{-3}$\, which, for a $\Lambda$CDM background, implies
\begin{equation}
K_{,X}-2G_{3,\phi}=6(\Omega_{m0}-1)(G_{4,X}+2XG_{4,XX}-G_{5,\phi})\,.\label{eq:lcdm2}
\end{equation}
Therefore, we see that the background evolution is exactly $\Lambda$CDM, the effective Newton constant
has the standard value (i.e. $Y=1$) but the anisotropic stress can be different from unity.

Let us finally give a specific case for which these conditions are satisfied. If, for instance, we take 
\begin{align}
K=\beta e^{mX}\,,\quad G_3=0\,,\quad
G_{4}=\alpha\frac{1+X^{n}}{1+X}\,,\quad G_5=0\,,
\end{align}
where $X$ is in units of $H_0^2$ and $\alpha$\,, $\beta$\,, $n$ and $m$ are constants, the eqs.  (\ref{a1}), (\ref{eq:newc1}), (\ref{omegaM}), (\ref{oM2})   and (\ref{eq:lcdm2})
can be simultaneously solved, provided $X$ is suitably chosen. 
For instance, if we choose $\alpha=-0.286,X=1.082,n=1.360$ and put $\Omega_m^0=1/3$
we obtain $\beta=-1.70,m=-0.251$ and $c_{T}\approx 0.8$ as in
our best fit, with the anisotropic stress
different from unity and equal to $\eta\approx1.55$. 
Although this is just a minimal toy model without any special physical significance,
it nevertheless shows that a MG model with the properties we have employed in this paper is indeed possible. 
More general cases in which, for instance, $G_3,G_5$ are not zero  and $\eta$ is time and scale dependent also exist. 
Many other examples in which $Y=1+\mathcal{O}\left(10^{-N}\right)$, where $N$ is large, can be constructed as well.

\bibliography{cmblensing}

\end{document}